# Changes in service distances of urban parks before and after the COVID-19 Pandemic :

# Applying a modified gravity model for Seoul Metropolitan Area

Da Won Oh[1] and In Kwon Park[2]

[1] [2]Graduate School of Environmental Studies, Seoul National University,

Seoul 151-742, Korea

## Author Note

**Abstract**

   COVID-19 significantly has changed the lifestyle in the urban areas. Urban parks reemerged as a savior of leisure activities and social joints under strict social-distancing measures. Physical activities such as strolling and jogging have relatively increased to avoid contact with others in confined indoor spaces such as shopping malls and gyms (Geng, Innes, Wu, & Wang, 2021). Also, there have been significant changes in the thresholds of service distances of urban parks: People have become more willing to visit parks in farther areas.

   This paper aims to examine the difference between before (May 2019, 31 days) and after the COVID-19 pandemic (May 2020, 31 days) by applying a gravity model. We examine variations in the service areas of urban parks, depnedig on the accessibility and design components of the park, using a dataset consisting of the park visitor's 'origin (home)' and 'destination (park)'. This   LBD (Location-based Big Data) provides the home location of the urban park visitor. The data was constructed by SK Telecom, using individual smartphone signal data on a daily basis.

   Adjusted coefficients are estimated by OLS(ordinary least squares) with cluster-robust standard errors to compare the difference between 2019 and 2020. Contrary to a common belief, the transit accessibility of the park plays a more significant role than the physical traits   of each parks. Accessibility itself determines a lot of the threshold distance of the park visit.

   While previous studies have identified the factors influencing the reaching distance of park services (Wang, Brown, & Liu, 2015), this study also attempts to determine how the effects of the factors have changed due to the COVID-19 pandemic. As proven in this study, the marginal effects of those factors vary before and after the pandemic. . By identifying the factors that determine the distance to visit in urban parks, it is possible to see which factors should be more focused on in planning small parks for residents in the neighborhood or large parks for more visitors from the entire region.

**Key words**

urban park, social infrastructure, COVID-19, service distance, location-based big data

## 1. Introduction

COVID-19 transformed urban life drastically – the hype of urbanites, such as mailing and cultural venues such as cinema and galleries, was doomed with the strict social distancing measure during the pandemic era. Nevertheless, parks won their original virtue again as a buffering place: urban parks reemerged as a savior of city dwellers' mental and physical sanity. Along with intense air pollution, urban parks nearly had become a relic of the past planning utopian orthodoxies-which were often considered as old-fashioned and a deprived sense of reality.

As the park became the only leisure space, the behavior of using the park also changed. Leisure activities in parks such as strolling and jogging have significantly increased (Geng, Innes, Wu, & Wang, 2021); however, no research has been done on whether people have visited more distant parks or whether the frequency of using parks near their home has increased.

Thus, we aim to verify whether the distance coefficient of park visiting trips has significantly changed or not. In order to do so, we used smartphone GPS data provided by the telecommunication company; this big data comprises of daily trips of individuals who live in Seoul allowed more microscopic approach that can quantify the change of park visiting travels between the pre-COVID and pandemic era.

This study focuses on the effect of park-wise physical factors such as the amount of green area, open space, and amenities (sports facilities and playgrounds). Furthermore, the transit accessibility of parks by subway and bus took into account automobile accessibility that represented road capacity and parking availability.

## 2. Review of Literature

### 2.1 Park use, leisure activity, and pandemic

There has been a great expectancy over behavioral change regarding leisure activities in urban parks. Since park visitation played a vital role in maintaining physical and mental well-being during the COVID-19 pandemic, multiple studies in various regions reported behavioral change. Restrictions on the social gathering, and the closure of indoor recreational places, such as malls and cafes, increased park visits globally (Geng, Innes, Wu, & Wang, 2021). As a result, the number of park visitors significantly increased with the population density of the cities, though usage varied considerably by the park in each city (Alizadehtazi, Tangtrakul, Woerdeman, Gussenhoven, Mostafavi, & Montalto, 2020).

During the pandemic, the beneficiaries of accessible urban parks proved their value by improving happiness: different types of urban parks are likely to affect residents' expressed happiness differently. As Cheng, Zhang, Wei, & Zhao (2021) analyzed, links between parks and residents' happiness become increasingly evident evermore.

Moreover, urban parks took a role as a successful barrier reducing the transmission rate of COVID-19. Johnson, Hordley, Greenwell, & Evans (2021) examined that higher park use was associated with reducing CoVID-19 transmission; rather than carrying out other indoor activities (e.g., visiting shops and workplaces), leisure time in parks significantly reduces the chance of getting infected. These results suggest that urban parks may successfully carry out their role as a buffering place during the pandemic.

## 2.2 Determinants of the threshold distance of the park visit

Initially, the distance of visiting urban parks is associated with the spatial configuration of parks and their distribution, representing the accessibility across neighborhood areas or local regions. The impact of the park-wise characteristics on threshold distance of park visits considered necessary in relevant studies since measuring park access is critical to evaluate the correlation of park use on physical activity (Kaczynski, Potwarka, & Saelens, 2008)

Park access, including park proximity to neighborhoods, park size, park safety, and park attractiveness regarding amenities&facility types, quality, and quantity, were used as the contents to measure park access in preliminary studies (Zhang, Lu, & Holt, 2011). Several studies examined the various factors that determine the park visting distance including visitor motivations, attractiveness, aesthetic features and size (Giles-Corti et al., 2012; Rossi, Byrne, & Pickering, 2015; Schipperijn et al., 2010; Sisneros-Kidd, D'Antonio, Monz, & Mitrovich, 2021)

However, there are limitations in distance decay parameters associated with spatial threshold distance measures. The distance decay or friction parameter (β) represents the magnitude of the parameter associated with distance. As Giles-Corti & Donovan (2002) denotes that the distance decay effects on spatial interaction processes or behaviors, reflected by this friction parameter, could be very context-specific (geographic settings such as urban, suburban, and rural areas) and could vary significantly among different activities, such as commuting and leisure trips. Also, the distance decay parameter could vary among the types of destinations within the same type of activities.

## 3. Data

The study area covers 31 days in May 2019 and May 2020. May was chosen because the variance of newly confirmed cases reported in South Korea is relatively stable in comparison with other months; During a period when the number of confirmed cases is suddenly increasing, it is difficult to observe a phenomenon that is maintained for a certain period because the drastic change might affect the behavioral choices. Thus,  it was considered as the most suitable timeline that can compare with 2019's data;

The data of 273 urban parks in Seoul were used in this study; the rest of the park data was unable to obtain because of the number of visitors' data collected from smartphone signals; a unit size larger than 10,000 square meters is needed minimum.

### 3.1 SKTelecom Mobile GPS Data

In order to compare and analyze the behavioral changes of park visitors before and after the spread of COVID-19, the Orientation(Home)-Destination(Park) data collected every day based on smartphone GPS was used. LBD (Location-based Big Data) makes it possible to determine the population who visited a particular park on a particular day. Due to privacy protection policy, the data do not provide individuals' personal information such as a home address, occupation, and another socioeconomic status; instead, South Korea's largest telecommunication company(SKTelecom) provides the number of daily park visitors from 25 autonomous districts in Seoul.

### 3.2 OSMNX

OSMnx is a Python package that obtains geospatial data from OpenStreetMap, which enables to model, project, visualize, and analyze real-world street networks; This package computes the origin-destination distance matrices to find shortest routes. In this study, the package is used to calculate the shortest path in urban networks that connects Orientation(Home)-Destination(Park) coordinates. The package obtained the distance variable in this study to obtain the minimum distance between two points more sophisticated than using Euclidean distance. The travel cost approach in this study includes more realistic measures by using the minimum distance from the residential neighborhood to the destination park calculated based on the distance between traffic nodes on the map.

## 4. Model: Mixed effects linear regression with cross classified data

The cross-classified model in equation (1) denotes the log-transformed number of visitors between the origin(neighborhood) and destination(park) $i$, $\beta_0$ is the model intercept, $xi$ denotes the value of the covariate for $i$, $\beta_1$ is the associated slope coefficient $u^{(2)}_{park(i)}$ and $u^{(3)}_{neigh(i)}$ denote the park and neighborhood random effects for that $i$, and $e_i$ denotes the $i$-level residual error. The subscripts park($i$) and neigh($i$) are 'classification functions' which return the park visited and neighborhood resided in by the visitor amount $i$, respectively, to the 2019 and 2020 models.

$$y_i = \beta_0 + \beta_1 x_i + u^{(2)}_{park(i)} + u^{(3)}_{neigh(i)} + e_i \quad \ldots (1)$$
$$u^{(2)}_{park(i)} \sim N(0, \sigma^2_{u(3)})$$
$$u^{(3)}_{neigh(i)} \sim N(0, \sigma^2_{u(2)})$$
$$e_i \sim N(0, \sigma^2_e)$$

In each model, the total number of $i$ is 6825 for 273 parks and 25 neighborhoods.

## 5. Results

| Model | 2019 (intercept only) | | 2019 (with predictors) | | | 2020 (intercept only) | | 2020 (with predictors) | | |
|---|---|---|---|---|---|---|---|---|---|---|
| **Fixed part** | Coeff. | S.E. | Coeff. | S.E. | | Coeff. | S.E. | Coeff. | S.E. | |
| **(Intercept)** | 9.47 | 0.12 | 27.56 | 0.51 | *** | 9.33 | 0.12 | 28.05 | 0.5 | *** |
| **log_distance (m)** | | | -2.09 | 0.01 | *** | | | -2.16 | 0.01 | *** |
| **log_park size (m²)** | | | 0.18 | 0.08 | ** | | | 0.2 | 0.08 | ** |
| **log_green areas (m²)** | | | -0.03 | 0.06 | | | | -0.04 | 0.06 | |
| **log_driveway areas (m²)** | | | -0.04 | 0.02 | ** | | | -0.04 | 0.02 | ** |
| **# subway stn. in 100m rad.** | | | 0.82 | 0.11 | *** | | | 0.75 | 0.11 | *** |
| **# bus stn. in 100m rad.** | | | 0.03 | 0.01 | ** | | | 0.03 | 0.01 | ** |
| **percentage of lake area (%)** | | | 0.03 | 0.03 | | | | 0.03 | 0.03 | |
| **percentage of plaza area (%)** | | | 0.51 | 0.26 | * | | | 0.48 | 0.25 | * |
| **presence of outdoor stage** | | | 0.23 | 0.24 | | | | 0.2 | 0.23 | |
| **presence of canteen** | | | 0.52 | 0.27 | * | | | 0.53 | 0.26 | ** |
| **AIC / BIC** | | | 11224.15 / 11319.74 | | | | | 11675.98 / 11771.58 | | |
| **Pseudo-R² (fixed effects)** | | | 0.60 | | | | | 0.61 | | |
| **Pseudo-R² (total)** | | | 0.93 | | | | | 0.93 | | |

*p>0.1, **p>0.05, ***p>0.01,

[Table 1] result (2019, 2020)

As the result shows, the coefficient of the distance variable decreased in 2020: this fact indicates that people tend to visit a park near their home rather than the parks located far away during the COVID-19 period. The global pandemic restricted the mobility of people regarding park visits in contrary to the increase of total mobility in urban parks. However, there were no severe mobility and park use restrictions during the COVID-19 in South Korea; therefore, the distance traveled among park visitors did not decrease significantly.

Moreover, variables representing the accessibility, such as the driveway areas, number of subway and bus stations in the perimeter, are the essential aspects of the park that attract more visitors in both 2019 and 2020. In addition, the park-wise physical features, such as the size of open spaces and presence of snack stalls.